\documentstyle[preprint,aps]{revtex}
\makeatletter
\outer\def\newclass #1#2%
    {%
    \edef \ObjectClassList {#1 \ObjectClassList}%
    \expandname\NewCount {#1counter}\global\csname #1counter\endcsname = 0
    \expandname\def     {#1def}{\SaveContents {#1}}%
    \expandname\def    {#1list}{\ListObjects {#1}}%
    \expandname\def  {#1num}##1{\LookUp {#1}##1 \Using {#1counter}#2\Label}%
    }%
\def\NewCount #1
    {%
    \global\advance \count10 by 1
    \global\countdef #1=\count10
    \wlog {\string#1=\string\count\the\count10}%
    }%
\def \expandname  #1#2{\expandafter #1\csname #2\endcsname}%
\def \ifundefined   #1{\expandname \ifx {#1}\relax}%
\let \ObjectClassList = \empty
\newclass {ref}\relax
\newcount \lastrefno	\lastrefno=-1
\newcount \refsequence	\refsequence=0
\def\Refnums#1%
    {%
    \RefRange#1-\end%
    }%
\def\lgycite#1
    {%
    [\RefRange#1-\end]%
    }%
\def\RefRange #1-#2\end%
    {%
    \RefNums #1,\end
    \def \temp {#2}%
    \ifx \temp\empty \else -\expandafter\RefRange \temp\end \fi
    }%
\def\RefNums #1,#2\end%
    {%
    \def \temp {#1}%
    \ifx \temp\empty \else \SkipSpace \temp#1\end\fi
    \ifx \temp\empty
	\ifcase \refsequence
	    \or\or ,\number\lastrefno
	    \else  -\number\lastrefno
	\fi
	\lastrefno = -1
	\refsequence = 0
    \else
	\edef\temp {{ref}\temp\space}%
	\expandafter \LookUp \temp \Using {refcounter}\relax
	\global\advance \lastrefno by 1
	\edef \temp {\number\lastrefno}%
	\ifx \Label\temp
	    \global\advance\refsequence by 1
	\else
	    \global\advance\lastrefno by -1
	    \ifcase \refsequence
		\or ,%
		\or ,\number\lastrefno,%
	    \else   -\number\lastrefno,%
	    \fi
	    \Label
	    \refsequence = 1
	    \ifx\Suffix\empty
		\lastrefno = \Label
	    \else
		\lastrefno = -1
	    \fi
	\fi
	\RefNums #2,\end
    \fi
    }%
\newif\ifSaveFile
\newif\ifnotskip
\newwrite\SaveFile
\def\savefilename {\jobname.sst}%
\def\OpenSaveFile   {\immediate\openout\SaveFile = \savefilename
		     \global\SaveFiletrue}%
\def\CloseSaveFile  {\immediate\closeout\SaveFile
		     \global\SaveFilefalse}%
\def\savestate%
    {%
    \expandafter \SaveCounters \ObjectClassList \end
    }%
\def\SaveCounters #1 #2\end%
    {%
    \def \temp {#1}
    \ifx \temp\empty \else
	\SaveObject {\number\csname #1counter\endcsname}{num}{#1counter}%
	\def \temp {#2}
	\ifx \temp\empty \else \SaveCounters #2\end \fi
    \fi
    }%
\def\Define #1#2#3%
    {%
    \def \temp {#1}
    \ifx \temp\Num	\global \csname #2\endcsname = #3 \else
    \ifx \temp\empty	\expandname\xdef    {#2}{#3}%
    \else		\expandname\xdef {#1_#2}{#3}%
    \fi \fi
    \SaveObject {#3}{#1}{#2}\iftrue
    }%
\def\Contents {\iftrue \SaveContents}%
\def\Num      {num}%
\let\IfDefine     = \Define
\let\IfContents   = \Contents
\def\Savesstout #1#2#3%
    {%
    \immediate\write\@sstout {\noexpand\IfDefine {#2}{#3}{#1}\noexpand\fi}%
    }
\def\SaveObject #1#2#3%
    {%
    \ifSaveFile \else \OpenSaveFile \fi
    \immediate\write\SaveFile {\noexpand\IfDefine {#2}{#3}{#1}\noexpand\fi}%
    \ifnum\SaveFile=\@sstout\else\Savesstout {#1}{#2}{#3}\fi
    }%
\def\SaveContents #1#2%
    {%
    \ifSaveFile \else \OpenSaveFile \fi
    \BreakLine
    \SaveLine {#1}{#2}%
    }%
\begingroup
    \catcode`\^^M=\active %
  \gdef\BreakLine %
    {%
    \begingroup %
    \catcode`\^^M=\active %
    \newlinechar=`\^^M %
    }%
  \gdef\SaveLine #1#2#3%
    {%
    \toks255={#3}%
    \immediate\write\SaveFile %
	{\noexpand\IfContents {#1}{#2}\LBrace\the\toks255\RBrace\noexpand\fi}%
    \endgroup %
    }%
\endgroup
\def\ListObjects #1#2%
    {%
    \ifSaveFile \CloseSaveFile \fi
    \def \ObjClass {#1}
    \let \IfContents = \GetContents
    \let \IfDefine   = \iffalse		\ReadFileList #2,\savefilename,\end
    \let \IfDefine   = \IfDoObject
    \let \IfContents = \iffalse		\input \savefilename
    \let \IfContents = \Contents
    \let \IfDefine   = \Define
    }%
\def\ReadFileList #1,#2\end%
    {%
    \def \temp {#1}%
    \ifx \temp\empty \else \SkipSpace \temp#1\end \fi
    \ifx \temp\empty \else \input #1 \fi
    \def \temp {#2}%
    \ifx \temp\empty \else \ReadFileList #2\end \fi
    }%
\def\GetContents #1#2%
    {%
    \notskipfalse
    \def \temp {#1}
    \ifx \ObjClass\temp \expandname
	\ifx {#1_#2}\relax \else \notskiptrue \fi
    \fi
    \ifnotskip \expandname \DefContents {#1_#2_}%
    }%
\def\DefContents #1#2{\toks255 = {#2} \xdef #1{\the\toks255}}%
\def\IfDoObject #1%
    {%
    \def \temp {#1}
    \ifx \ObjClass\temp \DoObject {#1}
    }%
\def\DoObject #1#2#3%
    {{%
    \expandname\ifx {#1fmt}\relax	\item {#3.}%
    \else				\csname #1fmt\endcsname {#3}\fi
    \expandname\ifx {#1_#2_}\relax	\expandname \CopyLabel {#2}%
    \else				\csname #1_#2_\endcsname \fi
    }}%
\def\CopyLabel #1{\expandafter \Gobble \string #1}

\def\LookUp #1#2 #3\Using#4#5%
    {%
    \expandname \ifx {#1_#2}\relax
	\global\advance \csname #4\endcsname by 1
	\expandname \xdef {#1_#2}{\number \csname #4\endcsname}%
	\let \temp = #5%
	\ifx \temp\relax \else \expandname \temp {#1_#2}\fi
	\expandname \SaveObject {#1_#2}{#1}{#2}%
    \fi
    \xdef \Label {\csname #1_#2\endcsname}%
    \gdef \Suffix {#3}%
    \ifx \Suffix\empty \else
	\xdef \Suffix {\expandafter\TrimSpace \Suffix\end}%
	\xdef \Label  {\Label\Suffix}%
    \fi
    }%

\def\Gobble	 #1{}%
\def\TrimSpace   #1 \end{#1}%
\def\SkipSpace   #1#2#3\end%
    {%
    \def \temp {#2}%
    \ifx \temp\space \SkipSpace #1#3\end
    \else \gdef #1{#2#3}\fi
    }%
\begingroup
\catcode`\<=1 \catcode`\{=12
\catcode`\>=2 \catcode`\}=12
\xdef\LBrace<{>%
\xdef\RBrace<}>%
\endgroup

\NewCount\RefOne	\NewCount\RefTwo
\NewCount\NumKnown	\NewCount\NumUnknown
\NewCount\NumChanged	\NewCount\LastNumChanged
\NewCount\IfOrdered	\NewCount\Done
\newcount\dummy 
\def\cite#1{\NumKnown=0\NumUnknown=0\Done=0\gdef\List{|<>#1,}%
    \loop\expandafter\reFsorT \List,\end\ifcase\Done\repeat%
    \ifcase\NumKnown\else%
        \NumChanged=0\LastNumChanged=0\Done=0%
        \loop\expandafter\sorTcitE \List\end\ifcase\Done\repeat%
    \fi\expandafter\citE \List\end}
\def\citE #1\end{\lgycite{#1}}
\def\reFsorT |#1<#2>#3,#4,\end{\def\temp{#3}%
    \ifx\temp\empty\Done=1
        \ifcase\NumKnown\gdef\List{#2}%
        \else%
	    \ifcase\NumUnknown\gdef\List{|<>#1,|}%
            \else\gdef\List{|<>#1,|#2}%
            \fi%
        \fi%
    \else\expandname \ifx {ref_#3}\relax%
             \ifcase\NumUnknown\gdef\List{|#1<#3>#4,}%
             \else\gdef\List{|#1<#2,#3>#4,}%
             \fi\global\advance\NumUnknown by 1%
         \else%
             \ifcase\NumKnown\gdef\List{|#3<#2>#4,}%
             \else\gdef\List{|#1,#3<#2>#4,}%
             \fi\global\advance\NumKnown by 1%
         \fi%
    \fi}
\def\storeone#1{\global\RefOne=#1}
\def\storetwo#1{\global\RefTwo=#1}
\def\Ordered#1#2{%
    \expandafter\storeone\csname ref_#1\endcsname%
    \expandafter\storetwo\csname ref_#2\endcsname%
    \gdef\temp{\number\RefOne}\gdef\temp{\number\RefTwo}
    \ifnum\RefOne<\RefTwo\IfOrdered=0\else\IfOrdered=1\fi%
    \gdef\temp{\number\IfOrdered}}
\def\sorTcitE |#1<#2>#3,#4|#5\end{%
    \def\tempa{#1}\def\tempb{#2}\def\tempc{#3}\def\tempd{#4}\def\tempe{#5}%
    \ifx\tempb\empty
        \ifx\tempc\empty
            \ifx\tempe\empty%
                \ifnum\NumChanged=\LastNumChanged\Done=1\gdef\List{#1}%
                \else\global\LastNumChanged=\NumChanged\gdef\List{|<>#1,|}%
                \fi%
            \else%
                \ifnum\NumChanged=\LastNumChanged\Done=1\gdef\List{#1,#5}%
                \else\global\LastNumChanged=\NumChanged\gdef\List{|<>#1,|#5}%
                \fi%
            \fi%
        \else
            \gdef\List{|<#3>#4,|#5}%
        \fi%
    \else
        \ifx\tempc\empty
            \ifx\tempa\empty\gdef\List{|#2<#3>#4,|#5}%
            \else\gdef\List{|#1,#2<#3>#4,|#5}%
            \fi%
        \else
            \Ordered{#2}{#3}\ifcase\IfOrdered%
                \ifx\tempa\empty\gdef\List{|#2<#3>#4,|#5}%
                \else\gdef\List{|#1,#2<#3>#4,|#5}%
                \fi%
            \else%
                \global\advance\NumChanged by 1%
                \ifx\tempa\empty\gdef\List{|#3<#2>#4,|#5}%
                \else\gdef\List{|#1,#3<#2>#4,|#5}%
                \fi%
            \fi%
        \fi%
    \fi%
    }
\let\@mainsst=\SaveFile
\let\@sstout=\@mainsst
\def\journal#1,#2,{{\it #1\/} {\bf #2},}
\def\jpc#1,{\journal J. Phys. Chem., #1,}
\def\jcp#1,{\journal J. Chem. Phys., #1,}
\def\jpp#1,{\journal J. Phys. (Paris), #1,}
\def\jpf#1,{\journal J. Phys. France, #1,}
\def\epl#1,{\journal Europhys. Lett., #1,}
\def\prl#1,{\journal Phys. Rev. Lett., #1,}
\def\pr#1,{\journal Phys. Rev., #1,}
\def\macromol#1,{\journal Macromolecules, #1,}

\def\deldot{\vec\nabla\kern -2pt\cdot}
\def\delcross{\vec\nabla\kern -2pt\times}

\def\\{\relax \ifmmode \backslash \else {\tt\char`\\}\fi }

\def\VEV#1{\left\langle #1\right\rangle}
\def\ave#1{\VEV{#1}}

\let\int=\intop         
\def\lsim{\mathrel{\mathpalette\@versim<}}
\def\gsim{\mathrel{\mathpalette\@versim>}}
\def\@versim#1#2{\vcenter{\offinterlineskip
	\ialign{$\m@th#1\hfil##\hfil$\crcr#2\crcr\sim\crcr } }}

\def\undertext #1{\vtop{\hbox{#1}\kern 1pt \hrule}}
\def\references{\section*{References\@mkboth
  {REFERENCES}{REFERENCES}}\list
  {[\arabic{enumi}]}{\settowidth\labelwidth{[99]}\leftmargin\labelwidth
    \advance\leftmargin\labelsep\itemsep=0pt\parsep=0pt
    \usecounter{enumi}}
    \def\newblock{\hskip .11em plus .33em minus .07em}
    \sloppy\clubpenalty4000\widowpenalty4000
    \sfcode`\.=1000\relax}

\makeatother
\begin{document}
\draft
\preprint{NSF ITP 95-24}
\def\noCite#1%
	{%
	\def \temp {#1}%
	\edef\temp {{ref}\temp\space}%
	\expandafter \LookUp \temp \Using {refcounter}\relax
	\global\advance \lastrefno by 1
	\edef \temp {\number\lastrefno}%
	}%
\makeatletter
\def\@oddhead{\hfill F.C. MacKintosh, J. K\"as, and P.A. Janmey}
\let\@evenhead\@oddhead
\global\@specialpagefalse
\makeatother
\tighten
\title{Elasticity of Semiflexible Biopolymer Networks}
\author{F.C. MacKintosh$^{1,2}$, J. K\"as$^{1,3}$, and P.A. Janmey$^{1,3}$ \\
$^1$Institute for Theoretical Physics, University of California, 
Santa Barbara, CA 93106-4030. \\
$^2$Department of Physics, University of Michigan, 
Ann Arbor MI 48109-1120.\\
$^3$Division of Experimental Medicine, Brigham and Women's Hospital, 
Harvard Medical School, Boston MA 02115.
}
\maketitle

\begin{abstract} 
We develop a model for gels and entangled solutions of semiflexible
biopolymers such as F-actin.  
Such networks play a crucial structural role in 
the cytoskeleton of cells.
We show that the rheologic
properties of these networks can result from nonclassical
rubber elasticity.
This model can explain a number of elastic properties of 
such networks {\em in vitro}, including
the concentration dependence of the storage modulus and yield strain.
\end{abstract}

A variety of semiflexible biopolymers and protein filaments
affect cell structure and function.
The most prevalent of these in eucaryotic cells is actin, which forms
the cytoskeletal rim \cite{Rev,Stossel93}. 
This actin cortex is a polymer gel that 
provides mechanical stability to cells, and plays a key role in cell motion.
Networks of actin and other protein filaments 
{\em in vitro} have been the subject of considerable recent interest
\cite{Janmey86,Elson88,
JanmeyCellBio91,Stossel93,Newman,Haskell94,Pollard,Janmey94,ActinReptation}, 
not only because of their structural role in cells, but also 
because of unusual viscoelastic properties of these networks.
Such protein filaments as actin are novel in
that they form viscoelastic networks, in which
$a\ll\xi\lsim\ell_p$, where $a$ is the size of a monomer, $\xi$
is the characteristic ``mesh'' size of the network, and $\ell_p$
is the persistence length of a chain.
In the case of actin, $\xi$ and $\ell_p$ are of order 1$\mu$m.
This, for instance, has permitted direct visualization of
polymer dynamics such as
reptation \cite{PGG,DE} by optical microscopy \cite{ActinReptation}.
Insight into the control of viscoelasticity in networks
of both natural and synthetic semiflexible polymers
in this intermediate regime 
is also important for the design of biocompatible materials.
For instance,
aqueous gels of stiff protein filaments or biocompatible polymers
have both structural and pharmaceutical applications.  
However, neither models of flexible-chain solutions nor models
of rigid-rod networks \cite{PGG,DE} are directly
applicable to such systems.
Here, we report a model for the elasticity 
of semiflexible polymer networks that can account for many of the observed 
properties of such networks {\em in vitro}.

Solutions of actin filaments {\em in vitro}
exhibit a polydisperse length distribution of about 
2 $\mu$m to 70 $\mu$m in length, with a mean
length of 22 $\mu$m \cite{Kaufmann}.
Between 36 $\mu$g/ml and 2 mg/ml F-actin 
forms entangled solutions. 
Many of the properties that are important for the function
of the actin cortex appear to arise from essential 
differences of F-actin networks from 
gels and concentrated solutions of flexible polymer chains.
We show that the rather large elastic moduli 
may not be generated by a classical entropic rubber elasticity,
since the persistence length $\ell_p$ of the filaments is comparable 
to or larger than the characteristic mesh 
size of the network $\xi$, as illustrated in
Fig.\ \ref{Picture}.
This semiflexible nature of the filaments can also 
explain both qualitatively and quantitatively 
a number of elastic properties of actin gels and solutions.
We shall focus primarily on 
actin networks, although our model is applicable to
other semiflexible polymers at intermediate concentrations.

Polymer gels and concentrated solutions are characterized by 
entangled chains.
Depending on preparation, these chains can be
chemically crosslinked.
For such a gel, 
the density of chemical crosslinks can be controlled by preparation.  
This density then determines the average distance between 
points along the chain that are effectively fixed by the 
bonds with the surrounding elastic network of other chains.
For an entangled solution, on the other hand, the viscoelastic properties
depend on transient 
entanglements of an individual chain with 
its neighbors \cite{PGG,DE}.
Despite the transient nature of these entanglements, 
over intermediate time scales of interest, 
the effect  
is much the same as that of chemical crosslinks, although the effective
degree of entanglement or the average length $L_e$ between entanglements
is more subtle \cite{NoteEntanglements}.
This intermediate regime is the ``rubber plateau'', for which the 
solution behaves as an elastic solid.
It is this regime that we address below.

Although many viscoelastic properties of actin and other
biopolymer networks resemble those 
of high molecular weight solutions of flexible polymer chains, the rubber 
plateau regime exhibits novel behavior.
Actin solutions, for instance, exhibit 
relatively high plateau moduli, 
of order 100 Pa or higher for actin monomer 
concentrations of order 1 mg/ml (i.e., for volume fractions
of order 0.1\%) \cite{Janmey94}.
The plateau modulus of actin networks also
exhibits significant strain hardening for modest strains. 
A rather small linear regime is observed:
e.g., in many cases they have a threshold strain as low as 
10\%, beyond which they lose their mechanical integrity. 
In the case of actin, this maximal strain also weakly decreases with 
increasing actin concentration \cite{Janmey88}.
As we show, this is a direct consequence of the 
intrinsic bending rigidity of biopolymers such as actin,
and is direct evidence of the inapplicability of the 
freely-jointed chain model for the concentrations of interest%
\cite{Nature90,Kasabstract,BendingModulus}.

We develop a model for densely crosslinked actin gels and entangled
solutions, in which the 
elastic properties arise from chains that are very nearly 
straight between entanglements, as illustrated in Fig.\ \ref{Picture}.
As we shall focus on the elastic rubber plateau modulus, 
we shall not distinguish 
between crosslinked gels and entangled solutions, except in 
so far as the entanglement lengths may differ.
We show that for an entangled solution, 
the plateau modulus scales with concentration $c_A$ as $G'\sim c_A^{11/5}$.
As shown in Fig.\ \ref{FigData} \noCite{Furukawa}
this is consistent with the measurements to date of the 
concentration dependence of $G'$ in the range
of $0.3-2.0$ mg/ml \cite{JanmeyCellBio91}.
For densely crosslinked gels, however, a somewhat stronger,
$G'\sim c_A^{5/2}$, dependence is predicted.

In our model for the linear elasticity in the 
plateau regime, we consider an ensemble of chain segments
of length $L_e$ (either between crosslinks or entanglement points),
which are embedded in a continuous medium that undergoes a uniform
shear deformation characterized by angle $\theta$.
The elastic response of the network results from 
the tension in such chain segments as a function of the extension, $L-L_0$,
where $L_0$ is the relaxed length.
When a semiflexible chain segment is stretched by a tension $\tau$, 
the energy per unit length of the chain depends on two effects: the
bending of the chain, and the work of contracting
against the applied tension.
The energy per unit length can be written \cite{NoteEnergy}
\begin{equation}
H={1\over2}\kappa\left(\nabla^2u\right)^2
 +{1\over2}\tau  \left(\nabla  u\right)^2, \label{Hamiltonian}
\end{equation}
where $\kappa$ is the chain 
bending modulus, and $u(x)$ describes the (transverse) deviation
of the chain away from a straight conformation along the $x$ axis.  
$\kappa$ is related to the persistence
length of the chain $\ell_p$ (the length over which
the chain appears straight in the presence of thermal 
undulations) by
$\ell_p\simeq\kappa/(kT)$. 
We let $L_\infty$ denote the full contour length of the chain
(i.e., for $\kappa=\infty$ or $\tau=\infty$).
We neglect the possibility of ``internal'' stretching of the chain:
i.e., the chain is assumed to have no longitudinal compliance.
Thus, for fixed contour length,
$L_\infty-L  \simeq {1\over 2}\int dx (\nabla u)^2$. 
At a given temperature and for a given tension $\tau$,
the transverse thermal fluctuations of $u$ determine the equilibrium
length $L$.
The chain conformation can be described 
by the Fourier series $u(x)=\sum_q u_q\sin(qx)$, where we include 
wavevectors $q={\pi\over L}, {2\pi\over L},\ldots$
consistent with fixed ends of the chain segment.
For the harmonic energy of 
Eq.\ (\protect{\ref{Hamiltonian}}),
the mean square amplitudes $\ave{u_q^2}$ 
can be calculated from the equipartition theorem, with the result that
\begin{equation}
L_\infty-L  \simeq {kT}\sum_q {1\over\kappa q^2 +\tau},
\end{equation}
where we have included both transverse polarizations of
$u$.
To linear order in applied tension $\tau$, 
the average end-to-end distance of the chain segment is 
$L\simeq L_\infty-kTL^2/(6\kappa)+kTL^4/(90\kappa^2)\tau$.
The second term represents the equilibrium contraction of the 
end-to-end distance at finite temperature.
The last term gives the linear relationship between the
applied tension and extension $\delta L$ of the chain segment
beyond its relaxed length.
For a small extension, 
the tension is given by \cite{NoteLargeTension}
\begin{equation}
\tau\sim {\kappa^2\over kTL^4}\delta L.\label{Tau}
\end{equation}

The above results for the behavior of individual chains 
can be used to estimate first the maximum shear strain
$\theta_{\rm max}$ that a network can withstand.
This will, in general, decrease with increasing concentration, since 
the entanglement length will then decrease.  This means that the 
fraction of the excess chain length in the form of thermal undulations
decreases, and hence there is less chain available to ``pull out'' under
the applied stress.
More precisely,
the relative extension of a segment of length $L_e$ between entanglements
is proportional to the strain $\theta$: 
$\delta L\sim \theta L_e$.
Considering the total excess length 
$L_\infty-L_0$ above,
the maximum strain for chain segments of length $L_e$
is given by
$\theta_{\rm max}\sim {kTL_e/\kappa}$.
Thus, the maximum strain is predicted to depend {\em linearly} on $L_e$.
Furthermore, this maximal strain decreases with increasing chain stiffness
(for the same entanglement length $L_e$).  This
is consistent with the observation that the yield strain does indeed
increase with increasing flexibility of the network:
networks of ADP actin, ATP actin, and vimentin show such a trend 
\cite{Kasabstract}.

For the modulus, $G'$, we use the relation above for the 
tension on an individual chain segment as a function of the shear strain
in the linear regime.  
For a network, we consider such a chain segment of length $L_e$ that is 
deformed 
by an amount given by $\delta L\sim \theta L_e$.
Within the linear regime (i.e., for small strain),
the tension in the chain segment is thus given by Eq.\ (\ref{Tau}).
Solutions and gels are characterized by a 
mesh size $\xi$ that describes the average spacing between chains
or the size of voids between filaments.
Along a plane parallel to the shear, there are $1/\xi^2$
chains per unit area \cite{DE}.
The stress $\sigma$ is therefore given by 
$\sigma\sim{\kappa^2/ (kT\xi^{2}L_e^3)}\theta$
in the linear regime.
Thus the modulus scales as
\begin{equation}
G'\sim{\kappa^2\over kT}\xi^{-2}L_e^{-3}.
\end{equation}
This is in contrast with the behavior of 
gels and networks of flexible chains, for which
$G'\sim kT/\xi^3$
\cite{PGG}.

Both the entanglement length $L_e$ and the mesh size $\xi$ 
decrease with increasing concentration of chains, 
although, unlike concentrated solutions of flexible chains, 
the scaling of these quantities with concentration
need not be the the same when $L_e\gsim\xi$ \cite{Rubinstein}.
The characteristic mesh size $\xi$ for a network of stiff chains is given by
$\xi \sim 1/\sqrt{ac_A}$, where $c_A$ is the concentration
of actin monomers of size $a$ \cite{TubeMeas}.
This is valid when the persistence length of the chains is longer than the 
mesh size $\xi$.
For a densely crosslinked gel, $\xi$ is also the typical distance 
between crosslinks, and therefore entanglement points: $L_e\simeq\xi$.
In this case, 
\begin{equation}
\theta_{\rm max}\sim {kT\xi\over\kappa}
\sim {kT\over\kappa}(ac_A)^{-1/2}
\end{equation}
and
\begin{equation}
G'\sim{\kappa^2\over kT}\xi^{-5}
\sim {\kappa^2\over kT}(ac_A)^{5/2}.\label{GPrimeGel}
\end{equation}

The precise dependence of the entanglement length on concentration in
a solution of semiflexible chains is more subtle.  
We expect that $L_e$ may become substantially larger than $\xi$ 
for $\xi\lsim\ell_p$,
since the transverse fluctuations of a semiflexible
chain are greatly reduced over distances comparable to or smaller
than the persistence length of the chain. 
For example, on the scale of the mesh size $\xi<\ell_p$, chains cannot 
form loops and knots \cite{Semenov,Tony,NoteTony}. 
We can estimate the entanglement length in the following way 
\cite{Semenov,Rubinstein}.  
{}From the above energy in Eq.\ (\ref{Hamiltonian}), 
the transverse fluctuations at temperature $T$ 
of a chain confined (entangled) at one end grow as
$\langle L_\perp^2\rangle\sim kTL^3/\kappa$, where $L$ is the 
distance from the entanglement.  
Thus, the fluctuating chain segments
of length $L_e$ between entanglements occupy a volume
$L_e\langle L_\perp^2\rangle\sim kTL_e^4/\kappa$. 
For a given concentration $c_A$, the probability of 
an intersection with another chain is of order unity for
$L_e\sim \left({\kappa/ kT}\right)^{1/5}\left(ac_A\right)^{-2/5}$,
which becomes larger than $\xi$ for $\xi\ll\ell_p$.
Thus
\begin{equation}
\theta_{\rm max} 
\sim \left({kT/\kappa}\right)^{4/5}(ac_A)^{-2/5}
\end{equation}
and
\begin{equation}
G'  \sim  \kappa\left({\kappa/kT}\right)^{2/5}(ac_A)^{11/5}.
\label{GPrimeNetwork}
\end{equation}

This model provides a consistent framework with which 
to understand the 
macroscopic viscoelasticity of biopolymer gels and solutions.
Based on the semiflexible nature of several
biopolymers, including F-actin, the model can explain
both the large storage moduli as well as the observed
strain hardening of  
networks at moderate to low strains \cite{CalcGandTheta},
a feature in contrast with the behavior of flexible polymer networks 
\cite{NoteRods}.  
For instance, at equal volume fractions, vimentin filaments, which are
approximately an order of magnitude less stiff than F-actin, form networks with
smaller shear moduli than F-actin, although
vimentin networks can withstand approximately 10 times larger strains 
than F-actin networks before rupturing.  
Experimental observations of shear moduli and yield strain
for varying actin concentration, as well as for modest changes in 
F-actin stiffness induced by binding of different 
nucleotides, are also in support of this model. 

This model 
makes several additional 
predictions that can be tested experimentally.  
First, as indicated above, 
for densely crosslinked gels, $G' \sim \kappa^2$.    
Since it is now possible to 
measure $\kappa$ directly for actin and some other biopolymers by 
video microscopy \cite{Kasabstract}, 
and there are a number of actin binding proteins and
metabolites that can alter filament stiffness under conditions where 
filament length is held constant, the viscoelastic parameters can be 
directly measured as a function of $\kappa$.   
Furthermore, the scaling behavior 
of entangled solutions and crosslinks gels as a function of concentration 
are predicted to differ.   
A third prediction is that the viscoelasticity of relatively dilute filament 
networks will be extremely sensitive to filament length even if the 
average filament length is much greater than the mesh size, and that this
dependence will be greatest for the stiffest polymers.  This is because for
semi-flexible filaments the entanglement length required for effects on
elasticity can be much greater than the mesh size, 
and this difference 
depends on $\kappa$.   
Therefore, subtle changes in filament length can have large
effects on viscoelasticity even when all filaments exhibit significant
overlap.   This feature may be one of the reasons the cytoskeletal actin
filaments in cells are under the tight control of proteins that regulate 
their length.

{The authors wish to thank
P.G. de Gennes, 
A. Maggs,
E. Sackmann,
and C. Schmidt
for helpful comments.
This work was supported in part by NSF Grant No. PHY94-07194.
F.C.M. was funded by NSF Grant No. DMR 92-57544.
J.K. was funded by a Forschungsstipendium of the DFG.
J.K. and P.J. were funded by NIH Grant AR38910.}

\def\journal#1, #2,{{\frenchspacing #1} {\bf #2},}
\def\j{\journal} 
\refdef{Aharoni}{S.M. Aharoni, \j Macromolecules, 25, 1510 (1992).}
\refdef{TubeMeas}{C.F. Schmidt, M. Baermann, G. Isenberg, and E. Sackmann, 
\j Macromolecules, 22, 3638 (1989).}
\refdef{Elson88}{E. Elson, \j Annu. Rev. Biophys. Biochem., 17, 397 (1988).}
\refdef{Stossel93}{T. Stossel, \j Science, 260, 1086 (1993).}
\refdef{Haskell94}{J. Haskell, et al., \j Biophys. J., 66, A196 (1994).}
\refdef{Marko}{C. Bustamante, J.F. Marko, E.D. Siggia, and S. Smith, 
\j Science, 265, 1599 (1994).}
\refdef{Janmey88}{P.A. Janmey, et al., \j Biochem., 27, 8218 (1988).} 
\refdef{EurBioPhys93}{R. Ruddies, W.H. Goldmann, G. Isenberg, and E. Sackmann, 
\j Eur. Biophys., 22, 309 (1993).}%
\refdef{Elson88}{E. Elson, \j Annu. Rev. Biophys. Biochem., 17, 397 (1988).}%
\refdef{KaesEPL}{J. K\"as, R. Strey, M. BarMann, and E. Sackmann, \j Europhys. 
Lett., 21, 865 (1993).}%
\refdef{DE}{M.Doi and S.F. Edwards, {\it Theory of Polymer Dynamics} 
(Oxford University Press, New York, 1986).}
\refdef{PGG}{P.G. de Gennes, {\it Scaling Concepts in Polymers Physics} 
(Cornell, Ithaca, 1979).}
\refdef{ActinReptation}{J. K\"as, H. Strey, and E. Sackmann, 
\j Nature, 368, 226 (1994).}
\refdef{Fixman}{M. Fixman and J. Kovac, \j J. Chem. Phys., 58, 1564 (1973).}
\refdef{Semenov}{A.N. Semenov, \j J. Chem. Soc. Faraday Trans. 2, 82, 317 
(1986).}
\refdef{Rubinstein}{R.H. Colby, M. Rubinstein, and J.L. Viovy, 
\j Macromolecules, 25, 996 (1992).}
\refdef{StiffnessDependanceOfTheta}{P.A. Janmey, unpublished results.}
\refdef{Kaufmann}{S. Kaufmann, J. K\"as, W.H. Goldmann, E. Sackmann, 
and G. Isenberg, \j FEBS Lett., 314, 203 (1992)}
\refdef{Furukawa}{R. Furukawa, R. Kundra, and M. Fechheimer, 
\j Biochem., 32, 12346 (1993).}
\refdef{Kasabstract}{J. K\"as, L.E. Laham, D.K. Finger, and P.A. Janmey, 
\j Mol. Biol. Cell, 5, 157a (1994).}
\refdef{Pollard}{D. Wachsstock, W. Schwarz, and T. Pollard, 
\j Biophys. J., 66, 205 (1994).}
\refdef{Newman}{J. Newman, et al., 
L. Gershman, L. Selden, H. Kinosian, J. Travis, and J. Estes,
\j Biophys. J., 64, 1559 (1993).}
\refdef{Haskell94}{J. Haskell, et al., \j Biophys. J., 66, A196 (1994).}
\refdef{Macro91}{(Macromolecules 1991, 24(11):3111; 
Journal of Cell Biology 1991, 113).}
\refdef{Janmey86}{P.A. Janmey, et al., \j J. Biol. Chem., 261, 8357 (1986);
P.A. Janmey, S. Hvidt, J. Lamb, and T.P. Stossel, 
\j Nature, 345, 89 (1990).}
\refdef{Nature90}{P.A. Janmey, et al., 
J. Lamb, and T.P. Stossel., 
\j Nature, 347, 95 (1990).} 
\refdef{JanmeyNature90}{P.A. Janmey, S. Hvidt, J. Lamb, and T.P. 
Stossel, \j Nature, 345, 89 (1990).}
\refdef{JanmeyCellBio91}{P.A. Janmey, et al., \j J. Cell Biol., 
113, 155 (1991).}
\refdef{Janmey94}{P.A. Janmey, et al., \j J. Bio. Chem., 269, 
32503 (1994).}%
\refdef{Janmeyrev}{P.A. Janmey and C. Chaponnier, \j Current 
Opinion in Cell Biology, 7, 111 (1995).}
\refdef{Rev}{E.L. Elson, \j Annu. Rev. Biophys. Biophys. Chem., 
17, 397-430 (1988).}
\refdef{Podgornik}{R. Podgornik and V.A. Parsegian, 
\j Macromolecules, 23, 2265 (1990).}

\refdef{NotePhysicalPicture}{%
These effects can be explained by the essential difference between
such networks and systems of flexible chains on the scale of the
mesh size: the persistence length of the chains can become comparable to
or larger than the mesh size.
However, despite the qualitative and quantitative differences from a
freely-jointed chain model when the persistence length of individual 
polymer chains is on the order of the mesh size, this model explains why
the scaling of the plateau modulus with concentration is
nevertheless similar to that of conventional gels.
In particular, previous reports indicate that the modulus of 
actin and fibrin networks increase with concentration as
$G'\sim c_A^x$, where $x\simeq2.1-2.2$ \cite{JanmeyCellBio91}.}

\refdef{NoteLargeTension}{We note that for large tension $\tau$, 
$(L_\infty-L)/L\sim kT\tan^{-1}\left(L/\pi\zeta\right)/(\tau\zeta)$,
where $\zeta=\sqrt{\kappa/\tau}$.
As the tension increases, $L$ approaches the full contour length.
In particular, this means that the tension $\tau$ must diverge as 
$\tau\sim1/(L_\infty-L)^2$ near full extension $L\simeq L_\infty$.
This is in contrast with the prediction of the 
freely-jointed chain model.
This was recently pointed out for DNA by
C. Bustamante, J.F. Marko, E.D. Siggia, and S. Smith, 
\j Science, 265, 1599 (1994).
See also M. Fixman and J. Kovac, \j J. Chem. Phys., 58, 1564 (1973).}
\refdef{Spit}{C.A. Vasconcellos, et al., \j Science, 263, 969 (1994).}
\refdef{Biomat}{N.A. Peppas and R. Langer, \j Science, 263, 1715 (1994).}
\refdef{Kuzuu}{M. Doi and N.Y. Kuzuu, \j J. Polym. Sci. Polym. Phys. Ed., 
18, 409 (1980).}

\refdef{NoteRods}{Strain hardening is also 
predicted by a theory for interpenetrating rigid rods
, but only at strains 
much larger than those experimentally observed for F-actin networks.
See M. Doi and N.Y. Kuzuu, \j J. Polym. Sci. Polym. Phys. Ed., 18, 409 (1980).}  

\refdef{BendingModulus}{For a concentration of $1.4$ mg/ml, 
the plateau modulus is also found to increase from $37$ Pa to 
$90$ Pa as a function of the bending modulus \refsequence=0\cite{Nature90},
which increases in this case from $0.7\times10^{-26}$ Jm 
to $2.0\times10^{-26}$ Jm 
\cite{Kasabstract}.}

\refdef{NoteEnergy}{The gradient $\nabla u$ characterizes the local
orientation of the chain relative to the $x$ axis.  Furthermore, the
gradient, $\nabla^2 u$, of this gives the local curvature of the 
chain.  For small curvatures of the chain, the energy is described
by the first term.  For small gradients of $u$, $(\nabla u)^2/2$ describes the 
local contraction of the chain along its axis.  As this ``crumpling''
occurs against the applied tension $\tau$, the second term gives the
energy due to tension.
Such a description 
is valid for a given chain segment for either a large bending stiffness 
or a sufficiently strong tension.}

\refdef{NoteOrientation}{Of course, 
the actual extension of any particular chain segment
depends on its
orientation relative to the shear direction.  However, the ensemble
average of this extension or compression of chain segments of
various orientations will only alter the numerical prefactor above.}

\refdef{NoteEntanglements}{Although 
elastically active constraints in solutions of long, 
overlapping semiflexible polymers need
not have the same structure as the tightly entwined 
entanglements in flexible polymer solutions or melts, 
in analogy with these systems we retain the term
and symbol $L_e$ for entanglement length.}

\refdef{NoteResultForTension}{The chain conformation can be described 
by the Fourier series $u(x)=\sum_q u_q\sin(qx)$, where we include 
wavevectors $q={\pi\over L}, {2\pi\over L},\ldots$
consistent with fixed ends of the chain segment.
For the harmonic energy of 
Eq.\ (\protect{\ref{Hamiltonian}}),
the mean square amplitudes $\ave{u_q^2}$ 
can be calculated from the equipartition theorem, with the result that
$L_\infty-L  \simeq {kT}\sum_q (\kappa q^2 +\tau)^{-1}$.}

\refdef{NoteConsequences}{%
This and other properties of actin and related biopolymers
can provide unique opportunities to study 
fundamental aspects of polymer networks.
For instance, the large bending stiffness of actin
filaments has allowed polymer dynamics associated with
reptation \cite{PGG,DE} to be studied by 
optical microscopy \cite{ActinReptation}.}

\refdef{NotePers}{Although it is often natural to describe the behavior of
semiflexible chains in terms of the persistence length,
in order to be clear regarding the dependence on stiffness and
temperature, we shall describe the model explicitly in terms of the 
mechanical bending stiffness $\kappa$ and the temperature $T$.}

\refdef{NoteTony}{It has been suggested in Ref.\ \refsequence=0\cite{Tony}, 
for instance, that at lower concentrations of order 0.3 mg/ml 
the entanglements described in
Ref.\ \cite{Semenov} may not lead to stretching of actin filaments that are
only a few microns long.  This results in smaller shear moduli and 
a weaker concentration dependence: $G'\sim c_A^{1.4}$.
Here, we have focussed on a limit valid for either high concentration
or high molecular weight (filament length).}

\refdef{Tony}{H. Isambert and A.C. Maggs, preprint.}

\refdef{CalcGandTheta}{Relative to the modulus of a flexible network,
$G'\sim kT\xi^{-3}$, the moduli of Eqs.\ (\ref{GPrimeGel}) 
and (\ref{GPrimeNetwork}) are enhanced by $(\ell_p/\xi)^2$ and 
$(\ell_p/\xi)^{7/5}$,
respectively.  Thus, for instance, for a network or gel with 
$\xi\simeq 0.3\mu$m and $\ell_p\gsim 2.0\mu$m, the modulus is 
enhanced by as much as two orders of magnitude.}

\begin{figure}[t]
\caption{Entangled network of semiflexible actin filaments.
(A) In physiological conditions, individual monomeric 
actin proteins (G-actin) polymerize to form double-stranded
helical filaments known as F-actin.  These filaments
exhibit a polydisperse length distribution of up to
70 $\mu$m in length.  The persistence length of these filaments
is of order 2 $\mu$m.
(B) A dense solution (1.0 mg/ml) of actin filaments, approximately
0.03\% of which have been labeled with rhodamine-phalloidin in order
to visualize them by florescence microscopy. The average distance 
$\xi$ between chains in this figure is approximately 0.3 $\mu$m.
Note the nearly straight conformation of the filaments on this scale.}
\label{Picture}
\end{figure}

\begin{figure}[t]
\caption{The plateau modulus $G'$ of actin networks as a function of
concentration in mg/ml \refsequence=0\protect{\cite{JanmeyCellBio91}}.  
The predicted scaling for entangled networks, 
from Eq.\ (\protect{\ref{GPrimeNetwork}}), is shown.
In this case, $G'\sim c_A^{11/5}$.
A nematic phase of actin filaments has been shown to 
form above a concentration of approximately 2 mg/ml
\refsequence=0\protect{\cite{Furukawa}}.
Our model is valid for the entangled isotropic regime
\refsequence=0\protect{\cite{Semenov}}, 
$(kT/\kappa)^2/a< c_A< kT/(\kappa a^2)$.}
\label{FigData}
\end{figure}

\end{document}